\documentclass[fleqn,10pt]{wlscirep}
\usepackage[utf8]{inputenc}
\usepackage[T1]{fontenc}
\usepackage{xurl}
\title{EpiGeoPop: A Tool for Developing Spatially Accurate Country-level Epidemiological Models}

\author[1,+]{Lara Herriott}
\author[1,+]{Henriette L. Capel}
\author[1,+]{Isaac Ellmen}
\author[1,+]{Nathan Schofield}
\author[1,+]{Jiayuan Zhu}
\author[4]{Ben Lambert}
\author[1,3]{David Gavaghan}
\author[1, 3]{Ioana Bouros}
\author[3]{Richard Creswell}
\author[1,2]{Kit Gallagher}

\affil[1]{SABS R³ CDT, University of Oxford, UK}
\affil[2]{Mathematical Institute, University of Oxford, UK}
\affil[3]{Department of Computer Science, University of Oxford, UK}
\affil[4]{Department of Statistics, University of Oxford, UK}

\affil[*]{gallagher@maths.ox.ac.uk}

\affil[+]{these authors contributed equally to this work}

\begin{abstract}
    Mathematical models play a crucial role in understanding the spread of infectious disease outbreaks and influencing policy decisions. These models aid pandemic preparedness by predicting outcomes under hypothetical scenarios and identifying weaknesses in existing frameworks. However, their accuracy, utility, and comparability are being scrutinized. Agent-based models (ABMs) have emerged as a valuable tool, capturing population heterogeneity and spatial effects, particularly when assessing intervention strategies. Here we present EpiGeoPop, a user-friendly tool for rapidly preparing spatially accurate population configurations of entire countries. EpiGeoPop helps to address the problem of complex and time-consuming model set up in ABMs, specifically improving the integration of spatial detail. We subsequently demonstrate the importance of accurate spatial detail in ABM simulations of disease outbreaks using Epiabm, an ABM based on Imperial College London's CovidSim with improved modularity, documentation and testing. Our investigation involves the interplay between population density, the implementation of spatial transmission, and realistic interventions implemented in Epiabm.
\end{abstract}
\begin{document}

\flushbottom
\maketitle
\thispagestyle{empty}

\section*{Introduction}

Mathematical models played a critical role over the course of the COVID-19 pandemic, aiding our understanding of the factors underlying disease spread and, consequently, strongly influencing policy decisions \cite{pagel2022role, wilk2022impact}. These models are a key tool for pandemic preparedness, allowing predictions under hypothetical scenarios to be made and their effects analysed \cite{mandal_imperfect_2022}. In addition to predicting the temporal evolution of epidemics, estimating the impact of a range of interventions (such as isolating infected cases and closing specific places) plays a significant role in guiding government policy \cite{ferguson_report_2020, brooks-pollock_modelling_2021}. The CovidSim model \cite{ferguson_report_2020} is one such mathematical model which informed the early pandemic response in the UK and US by predicting the impact of different government interventions \cite{adam2020special}. Given the ongoing risk of future pandemics \cite{madhav2018pandemics}, improving the accuracy and usability of such models is an essential step towards pandemic preparedness. 

Agent-based models (ABMs) are a key modelling framework, alongside differential equation-based models \cite{Kermack_1927}. ABMs represent members of a population individually and thus natively handle population heterogeneity \cite{ferguson_strategies_2006, hunter_comparison_2018}. They allow detailed and informative predictions to be obtained in settings where inter-individual variability in traits, behaviour and interactions is important \cite{hunter_comparison_2018}. In contrast, differential equation models describe the spread of a disease through a system of equations modelling continuous state variables, resulting in a deterministic model that averages individual traits across a population \cite{hethcote_mathematics_2000, VANDERVEGT2022108824}. To represent an epidemic, differential equation models contain compartments representing the various stages of disease transmission, such as susceptible, exposed \cite{PHE_model, Keeling_2021, Moore_2022, Roche}, infected (possibly with multiple degrees of severity \cite{Roche}), and recovered. 

While differential equation models can be adapted to incorporate population heterogeneity, this involves adding additional compartments such as for age structure \cite{PHE_model, Keeling_2021, VANDERVEGT2022108824} or within-household infection pathways \cite{Keeling_2021}. However, increasing the number of compartments requires a corresponding increase in the number of equations in the model, which effectively places an upper limit on the amount of detail which can realistically be included within this modelling framework. As a consequence, very few models implement multiple levels of population heterogeneity simultaneously. Furthermore, by breaking the population down into compartments the number of individuals represented by each compartment reduces. When predicting transmission among small groups of individuals stochastic effects become important yet cannot be accounted for with continuous differential equation models. On the other hand, such information can more easily be incorporated in an ABM by adding descriptors to the individuals and then using these descriptors to scale or alter existing behaviours. 

The utility of this approach is apparent when assessing the impact of a range of interventions on mitigating the spread of disease, which is key to informing government policy. Many non-pharmaceutical interventions are very difficult to represent faithfully in deterministic equation models, and other highly targeted measures are beyond the scope of these models. School closure, for example, can only be captured in compartment-based models by adjusting the parameters governing the behaviour of school-aged children \cite{fung_modeling_2015} and is therefore only captured approximately and indirectly. In contrast, school closure can be implemented in an ABM by preventing virtual children from attending virtual schools \cite{ rice_effect_2020, hunter_using_2021}. Furthermore, additional complexities and knock-on effects of such an intervention, such as decreased workplace visits among parental groups, could also be captured \cite{litvinova_reactive_2019}. As such, modelling the effect of interventions is a major advantage of ABMs. However, including such individual-level detail in these models comes at a cost, either in time spent sourcing the relevant spatial and demographic data, or through the additional assumptions which must be made \cite{hunter_hybrid_2020, hunter_taxonomy_2017, noauthor_neil_2005}. In short, ABMs are `data hungry' \cite{adam_special_2020}.

Given their ability to capture substantial population heterogeneity, ABMs are also well-suited to considering the effect of spatial heterogeneities on disease spread. While few studies had explicitly considered the impact of population density on the large-scale spread of infectious diseases prior to the COVID-19 pandemic \cite{hu2013scaling, hamidi_does_2020}, the collection of huge amounts of infection data globally over the past three years have allowed this topic to be explored in greater detail. For example, understanding how the distribution of population density shapes pandemics is crucial for informing containment and mitigation policies and determining the effectiveness of localised restrictions \cite{fan_effect_2021, karatayev_local_2020}. In this way, an understanding of the effects of spatial patterns could improve pandemic preparedness.

In the present work, we begin to address the difficulty incorporating real-world data into  spatially accurate ABMs by presenting a user-friendly tool, EpiGeoPop for configuring spatially-detailed ABMs easily and rapidly. Previous tools and models have been developed to streamline the construction of ABMs \cite{bisset_epifast_2009, grefenstette_fred_2013, yeom_overcoming_2014, bhattacharya_data-driven_2023}. However, to the best of our knowledge, these pipelines process data for a single country only and are intrinsically linked to specific epidemiological models. In contrast, EpiGeoPop utilises spatially segregated population data for geographical regions across the globe. This is compiled into human-readable input files. In addition, EpiGeoPop generates population-specific parameter estimates from published data. EpiGeoPop can be used as a bolt-on to an existing epidemiological model to help speed up population configuration. 

We demonstrate the utility of EpiGeoPop, and the importance of incorporating accurate spatial detail into epidemiological models, by running example simulations on a national scale. For this we use the epidemiological model Epiabm, an alternative implementation of CovidSim \cite{ferguson_report_2020} which emphasizes modularity, documentation, and good software engineering practices \cite{gallagher_epidemiological_2022}. Epiabm and EpiGeoPop were developed as part of the first-year training in software engineering for PhD students at the EPSRC CDT in Sustainable Approaches to Biomedical Science: Responsible and Reproducible Research (SABS:R\textsuperscript{3}) CDT at the University of Oxford. 

\section*{Methods}

Here, we describe EpiGeoPop and discuss our improvements and extensions to Epiabm. To extend Epiabm for the present work we have continued the development of the Python back-end of this software. The original release of Epiabm did not implement pharmaceutical or non-pharmaceutical interventions, and used a simplistic implementation of spatially-mediated infections. Our extensions to Epiabm include both forms of intervention and an improved method of weighting spatial infections. The code for both EpiGeoPop and Epiabm are available open source at \url{https://github.com/SABS-R3-Epidemiology/EpiGeoPop} and \url{https://github.com/SABS-R3-Epidemiology/epiabm}.

\subsection*{EpiGeoPop}

\begin{figure}
\centering
\includegraphics[width=0.9\textwidth]{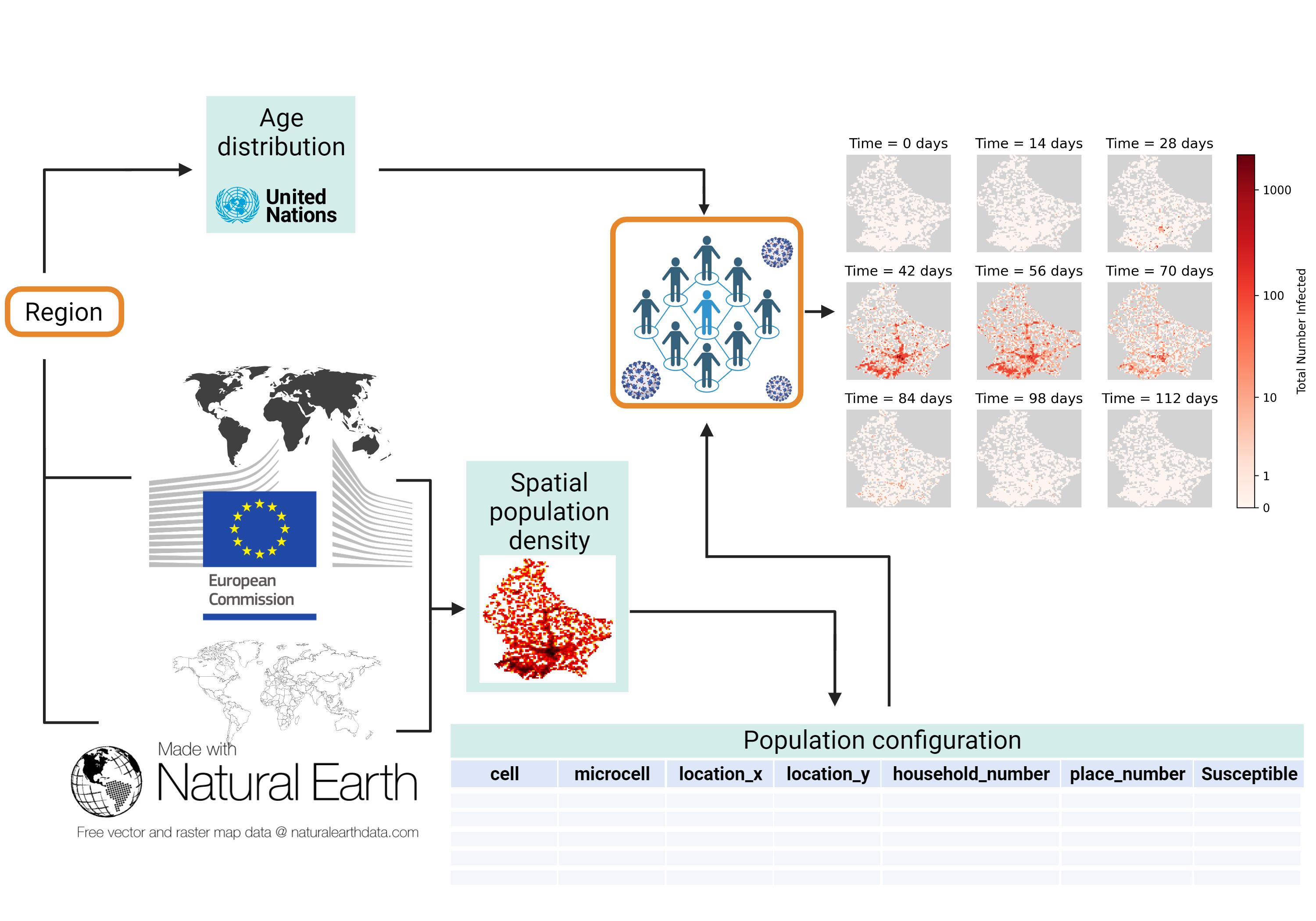}
\caption{EpiGeoPop workflow. EpiGeoPop uses global population density data to generate population configuration files for any country, province, or city of interest. In addition, age distributions are generated on a country-level basis. The choice of region and epidemiological model, highlighted in orange, represent the elements of the workflow determined by the user. As long as the overall format of the output file from the model matches that of the input file generated by EpiGeoPop, the visualisation module can be used. This allows geographically accurate visual summaries of disease dynamics to be generated in the form of GIFs. Created with BioRender.com.}
\label{fig:workflow}
\end{figure}

A pre-requisite for spatially accurate epidemiological modelling is the integration of spatial data into the modelling framework. To simplify these initial stages of model set up, we develop a workflow named EpiGeoPop (Figure \ref{fig:workflow}) which constructs a geographically accurate density map from global population density data \cite{pop_density} and converts this into human-readable population configuration file.
Our tool allows this input file to be generated for any country or region of interest for which the border and population density data is available in the Natural Earth  (\url{https://www.naturalearthdata.com/downloads/10m-cultural-vectors/}) and European Commission, Joint Research Centre Global Human Settlement Layer \cite{pop_density} databases, respectively. We use the 2015 version of the population density data set, available at: \url{{https://jeodpp.jrc.ec.europa.eu/ftp/jrc-opendata/GHSL/GHS_POP_MT_GLOBE_R2019A/GHS_POP_E2015_GLOBE_R2019A_4326_30ss/V1-0/}} - this includes regions on the scale of cities, provinces and complete countries.

\subsubsection*{The EpiGeoPop workflow}

For ease of use, we present a complete workflow as a user-friendly Snakemake pipeline \cite{snakemake} implemented in Python. The region of interest is broken down into subsections which we label `cells’, as in Epiabm. The data set automatically used by EpiGeoPop has a resolution of 30 arc seconds (approximately 1 km\textsuperscript{2}), which corresponds to the size of each `cell'. Subsequently, each `cell’ is split into a grid consisting of `microcells’, representing smaller geographical regions within which households, workplaces, and schools exist. In the simulations presented herein, we use a grid size of 3x3 totalling nine microcells per cell.

EpiGeoPop also generates country-specific age distributions, taken from the UN 2022 Revision of World Population Prospects \cite{age_dist}. These are processed to represent the proportion of individuals in 5 year age groups from 0-80, with a final group of over 80 year olds. 

Details on how to execute the work flow are provided in the Supplementary Information, Section 1.1.

\subsubsection*{Visualising pandemic spread}

We also provide code for generating time-series animations of pandemic spread across the region of interest (for example Supplementary Videos 1 and 2). This code is run as a single Python script which creates a GIF composed of the infection heatmaps for a location at each time point. To utilise this script, the output data from the epidemiological simulation should be provided in the format matching that generated by the Snakemake workflow (details are provided at \url{https://github.com/SABS-R3-Epidemiology/EpiGeoPop/#generating-animations}).

\subsection*{Epiabm}

We utilise Epiabm for all epidemiological simulations, extending this software from the previously published version. Our extended version weights spatial transmission more accurately, and implements a suite of both pharmaceutical and non-pharmaceutical interventions. 

\subsubsection*{Improved weighting of spatial transmission}
Spatial transmission across cells involves the random selection of one cell for each person a given infector should infect. In both CovidSim and, consequently, the initial version of Epiabm, this cell was selected with probability inversely proportional to its distance from the infector's cell:

\begin{equation}
    p_{cell} \propto \frac{1}{d},
\end{equation} where $p_{cell}$ is the probability of a cell being selected and $d$ is the distance between this cell and the infector's cell.

However, by equally weighting cells with different population sizes the selection is biased towards individuals in the least populated cells. We therefore adjusted the spatial infection mechanism by selecting this cell with probability inversely proportional to the distance from the infector's cell, and proportional to the population of the infectee's cell:

\begin{equation}
    p_{cell} \propto \frac{n}{d},
\end{equation} where $n$ is the population of the infectee's cell.

Because the probability of selecting an individual randomly from a cell is $1/n$, the probability of selecting an individual ($p_{ind}$) was originally:

\begin{equation}
    p_{ind} \propto \frac{1}{d} \times \frac{1}{n} = \frac{1}{d \times n},
\end{equation} whereas under the adjusted probability of selecting an individual is:

\begin{equation}
    p_{ind} \propto \frac{n}{d} \times \frac{1}{n} = \frac{1}{d}.
\end{equation}

In this way, the original selection meant individuals in highly populated cells such as dense urban centres were less likely to be infected. The new selection only considers the distance between the infector and all potential infectees, eliminating the bias related to local cell population, which we posit is a more realistic model of human contacts.

\subsubsection*{Non-pharmaceutical interventions}
We additionally implemented the full suite of non-pharmaceutical interventions present in CovidSim \cite{ferguson_report_2020}: case isolation, household quarantine, place closure, and social distancing. In our implementation of these interventions, we closely follow that used in CovidSim. Interventions can be activated based on time and/or a case threshold, capturing some of the flexibility present in CovidSim. Multiple interventions can be active at the same time, and the same intervention can be implemented over different time periods with differing severity. All non-pharmaceutical interventions influence house, place, and spatial infections. When case isolation is active, symptomatic or positive-tested individuals will isolate within their household for a given period (for a description of the testing functionality please see below under the section `Pharmaceutical interventions'). Household quarantine is built upon case isolation and restricts members of the household of the isolated individual to also stay at home for a defined number of days. Place closure allows for the closure of specific place types to be simulated. Social distancing has two types, enhanced or non-enhanced, with age-dependent probability of following enhanced social distancing which more strongly reduces the likelihood of house, place, and spatial infections.

Simulations including these interventions have the expected effect of delaying and reducing the peak of infections (Figure S1; see also \url{https://github.com/SABS-R3-Epidemiology/epiabm/tree/main/python_examples/intervention_example}).

\subsubsection*{Pharmaceutical interventions}
In addition, we implement two pharmaceutical interventions: mass vaccination and infection testing. Mass vaccination can be prioritised according to individual characteristics, and the protection afforded by vaccination can be parameterised by the user. Infection testing includes two testing streams for which parameters including capacity, sensitivity, selectivity and the type of user can be set. These two streams can therefore be adjusted to represent different testing modalities, such as PCR and lateral flow tests.

\subsubsection*{International travel}
Finally, we develop a simple travel framework to capture international travel. Through this functionality, border closure and isolation of international arrivals (either within an existing household or in a separate hotel) can be activated as an additional intervention.

\section*{Results}

We present simulations which show the importance of including spatial population density detail in epidemiological models. In addition, we investigate how interventions differentially impact different density regions. These simulations also demonstrate possible applications of both EpiGeoPop and Epiabm for investigating epidemic dynamics.

\subsection*{Population density influences epidemic dynamics}

\begin{figure}
\centering
\includegraphics[width=0.5\textwidth]{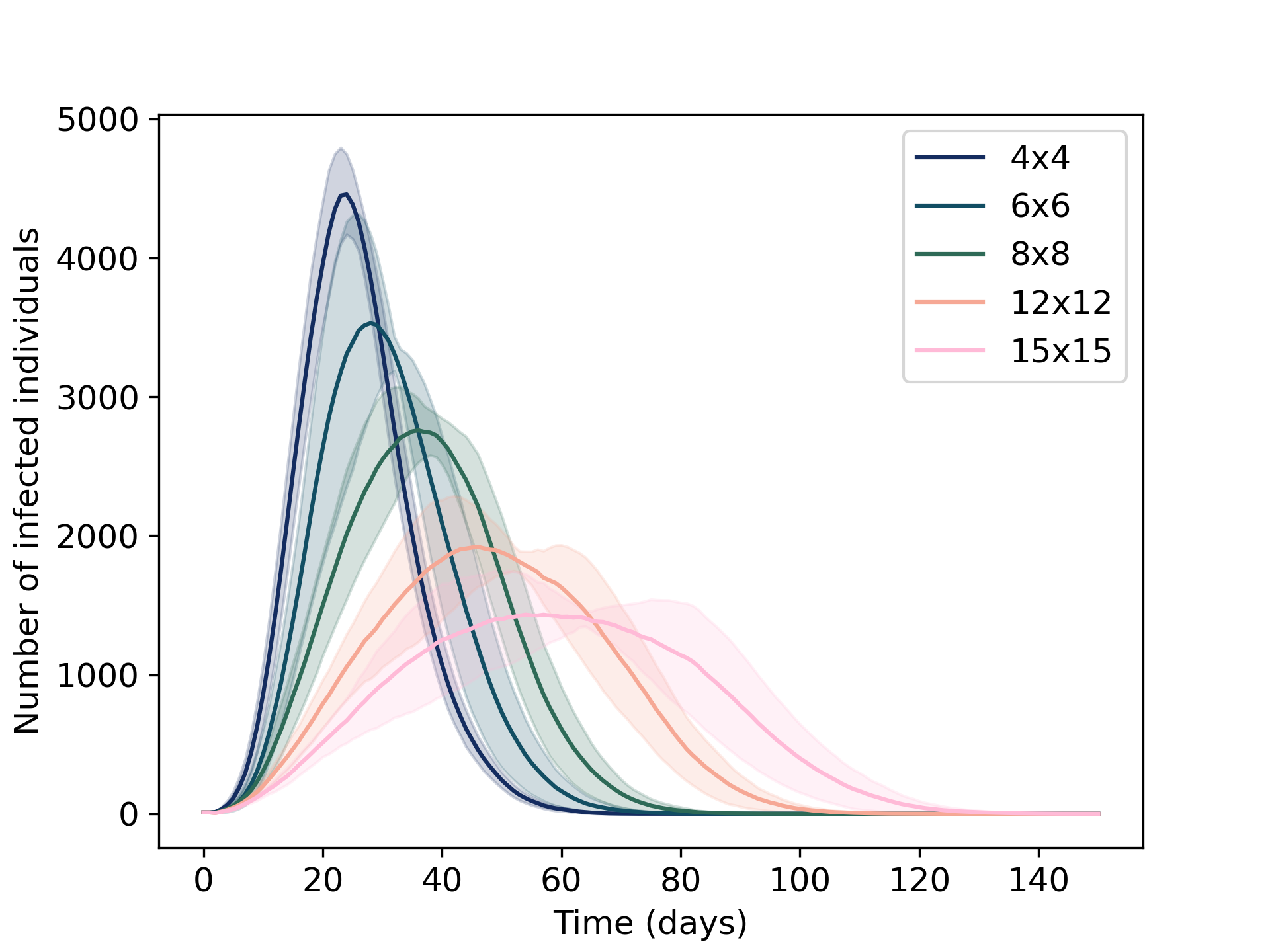}
\caption{Lower density populations experience delayed and lower peaks of infections. Populations of 10,000 individuals are spread approximately uniformly over square grids of dimensions 4x4, 6x6, 8x8, 12x12, and 15x15. As such, the 4x4 model population represented the highest density, while the 15x15 is the lowest density. Lines indicate the mean over 10 simulations, while the shaded areas represent the standard deviations.}
\label{fig:toy_populations}
\end{figure}

Simulations for five approximately uniform populations with differing population density reveal a pattern of lower and later peaks in lower density populations. In our highest density population (4x4 grid), the number of infections reaches an peak 30 days earlier and at a value more than three times greater than that in the lowest density population (15x15 grid) (Figure \ref{fig:toy_populations}). These results indicate the substantial influence population density can have on the spread of infections and motivate the importance of accurately replicating population density when simulating disease spread in real countries with heterogeneous population distributions.

% Understanding how the epidemic may spread through different regions of the country could also be important for developing strategies to target interventions or support to where they are most needed, or where they are most likely to be effective.

% \begin{figure}[H]
% \centering
% \includegraphics[width=\textwidth]{Images/voronoi_grid_log_img_post_change.png}
% \caption{Comparison of short \emph{versus} country-wide infection radii. A) Map of disease spread with an infection radius spanning approximately 5 cells. B) Map of disease spread with an infection radius encompassing the entire country. \textbf{NB:} comparison plot showing circular wave to be added.}
% \label{fig:circular_wave}
% \end{figure}

Using EpiGeoPop, we constructed a spatially accurate model of Luxembourg. We selected Luxembourg for investigative purposes since its population density varies widely across the country, while its population size ($n=569,335$, as of 2015) remains sufficiently small for simulations to be run locally in a reasonable time ($\sim45$ minutes). Details of all simulations run for Luxembourg can be found in the Supplementary Information (Section 1.3).

\begin{figure}
\centering
\includegraphics[width=0.9\textwidth]{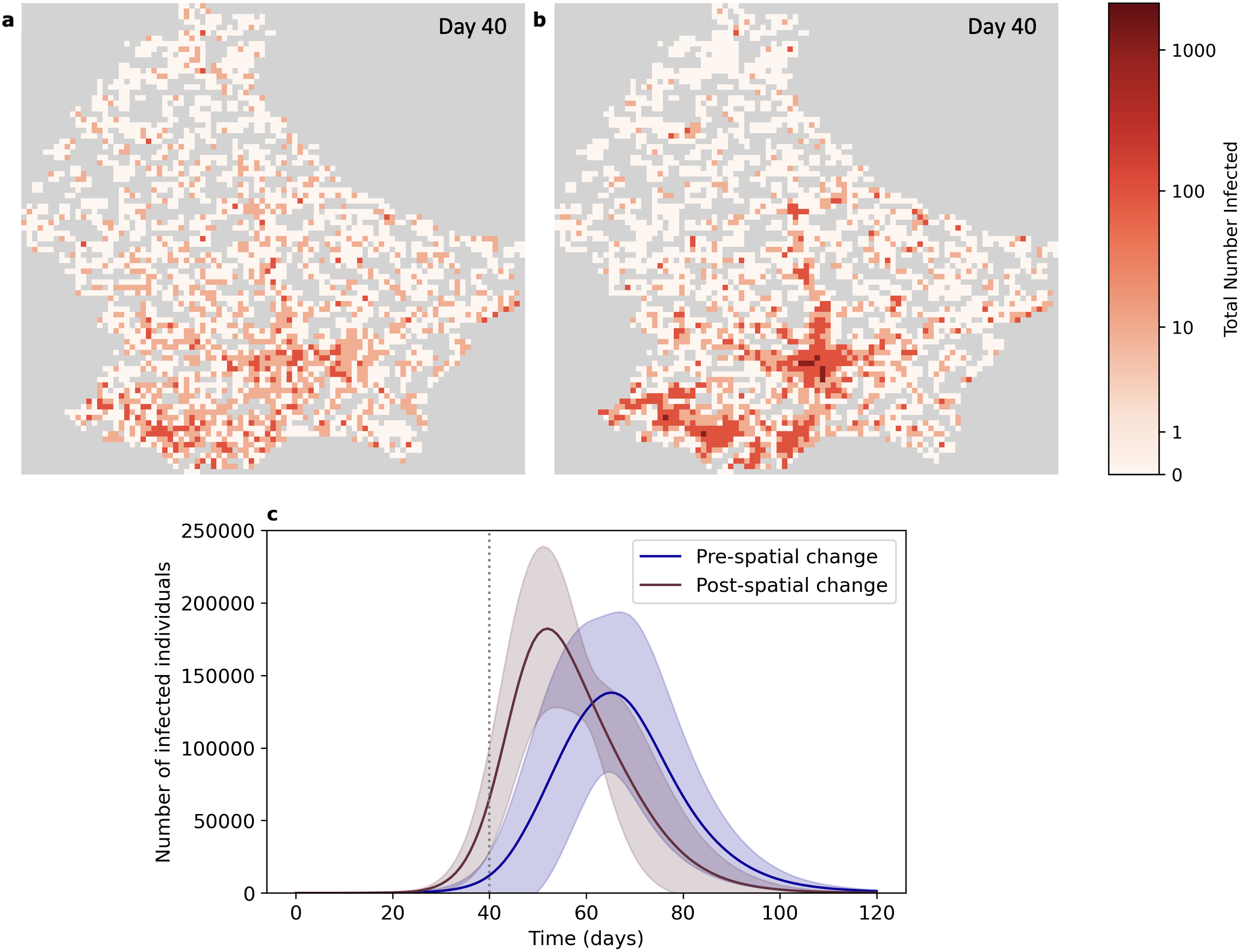}
\caption{Disease transmission in Luxembourg before and after changing the spatial transmission weightings. \textbf{(a)} Snapshot of disease transmission at day 40 with spatial infections guided by distance only. \textbf{(b)} Snapshot of disease transmission at day 40 with spatial infections guided by distance and population. In both snapshots darker colours indicate higher numbers of infections. \textbf{(c)} Total number of infections over time before (blue) and after (red) the change to the spatial transmission weightings. Results are shown as the mean ± standard deviation over 10 repetitions of each simulation. The grey dashed line indicates day 40, corresponding to the time of the above snapshots.}
\label{fig:spatial_change}
\end{figure}

As a result of our change in the spatial spread of disease, such that new infections are now guided by both proximity to the infector and population density, the epidemic is now biased towards entering large cities earlier: Figures \ref{fig:spatial_change}a, \ref{fig:spatial_change}b display snapshots of the distribution of infections at day 40.  Comparing these snapshots to the population density map in Figure \ref{fig:urban_rural_all}a, it can be seen that the distribution of infections at day 40 with the spatial change aligns with the population density distribution. Time-series animations of disease spread before and after the change in the spatial spread of disease can be found in Supplementary Videos 1 and 2. In addition, accounting for population density in spatial transmission results in a substantially shorter and sharper wave of infections, falling 13 days earlier and peaking at a 32\% higher value, which corresponds to 44,180 additional cases at the peak (Figure \ref{fig:spatial_change}c). 

%In the former case, infections peaked at 182,311 on day 52, compared to a peak of 138,131 at day 65 in the latter case. 

% By guiding new spatial infections by both distance and population density, our model more accurately captures real demographic patterns, such as the greater movement of people towards and between large population centers, and the longer distances travelled by more rural individuals to access resources in regions of higher population density (Chart 34, National Travel Survey; Figure 3, Travel to work).

\subsection*{Epidemic dynamics differ in different density regions}

\begin{figure}
\centering
\includegraphics[width=0.9\textwidth]{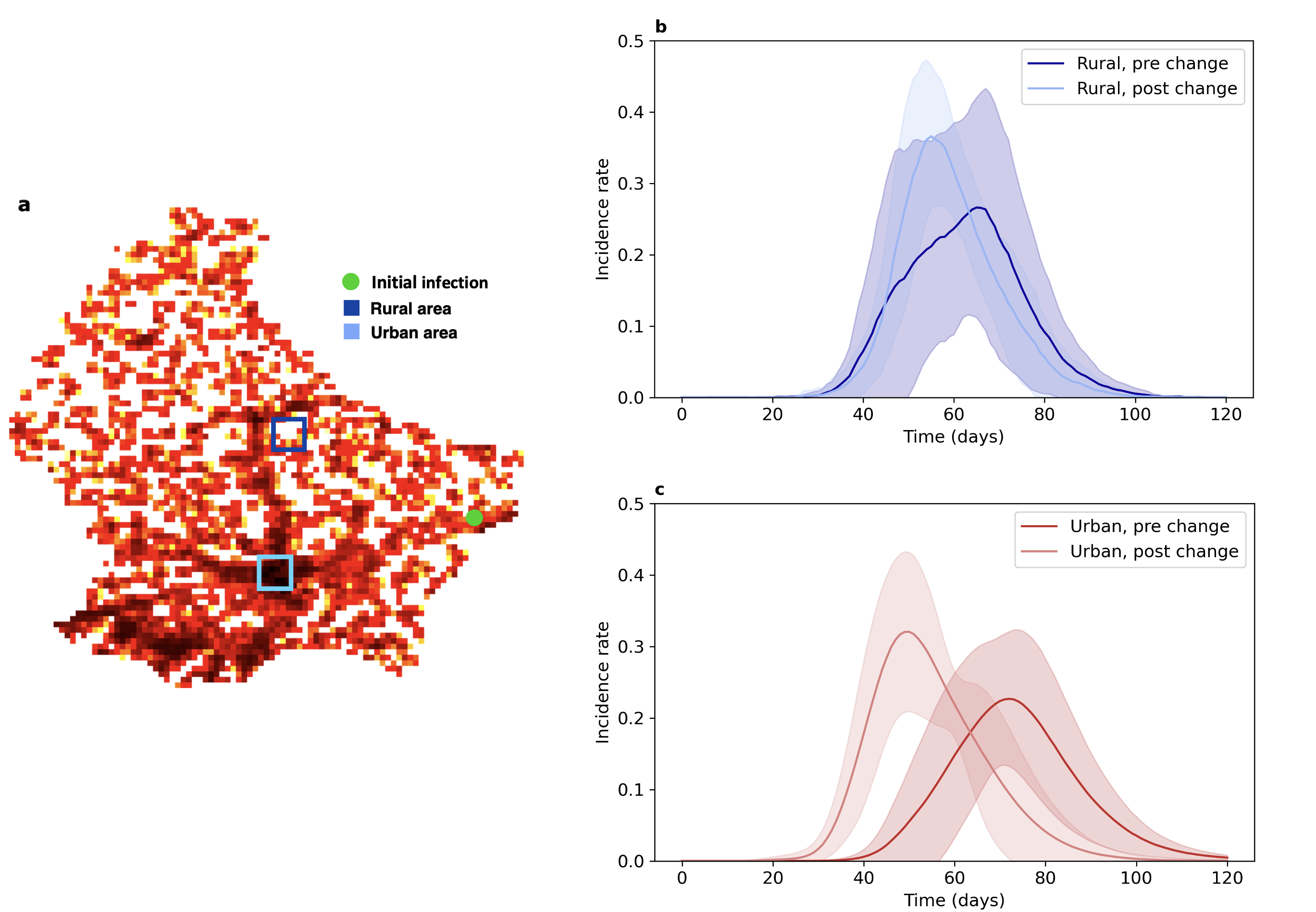}
\caption{Difference in the spread of disease in Luxembourg in rural and urban populations. \textbf{(a)} Map highlighting the locations of the urban and rural regions selected for analysis. The urban region is highlighted in light blue and the rural region in dark blue. The location of the initial infections is indicated by the green circle. \textbf{(b)} Incidence rate over the course of the simulation for the selected urban and rural regions both before (dark red and dark blue, respectively) and after (light red and light blue, respectively) the change to the spatial weighting. Results are shown as the mean ± standard deviation over 10 repetitions of each simulation. The incidence rate is calculated separately for each region.}
\label{fig:urban_rural_all}
\end{figure}

To further elucidate how population density influences disease spread, we compared the shape of the epidemic curve in an urban region, Luxembourg City, with that in the more rural commune of Nommern (Figure \ref{fig:urban_rural_all}a). These locations were selected since they are approximately equidistant from the location selected to seed the initial infections. 

Under both the old and the new mechanisms of spatial transmission, the lower density population experienced a wave of infections over similar timescales (Figure \ref{fig:urban_rural_all}b). However, the higher density region experienced a much earlier wave of infections after the change to the spatial transmission weightings. As indicated in the infection heatmaps above (Figure \ref{fig:spatial_change}a, \ref{fig:spatial_change}b), when spatial infections were guided by distance and population the epidemic enters urban regions earlier. Interestingly, these results demonstrate that this also translates to an increase in the peak number of infections in a densely populated region. Under the new spatial transmission weightings, the rural region exhibited a slightly later peak of infections than that observed for the higher density population (Figure \ref{fig:urban_rural_all}b).

% This delayed peak of infections in lower density populations compared to higher density ones could be considered to buy policy makers more time to make decisions surrounding interventions, such that when they are implemented they come into effect at an earlier stage of the wave of infections when the total case burden is lower, and so would be more effective.   

\subsection*{Rural and urban environments are similarly impacted by interventions}

Based on these differences, we consider whether the impact of interventions differed between higher and lower density regions. We applied a combination of interventions to Luxembourg, comprising case isolation and household quarantine for the duration of the 150-day simulation and social distancing from day 49.

\begin{figure}
\centering
\includegraphics[width=0.9\textwidth]{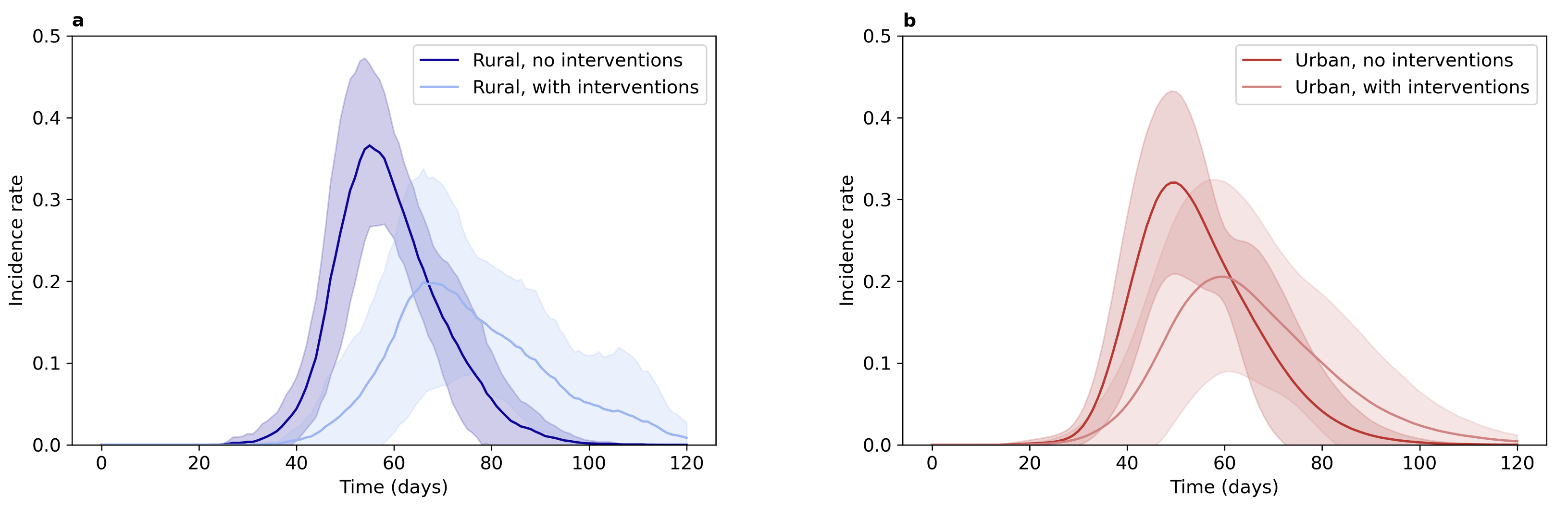}
\caption{Incidence rate over the course of the simulation for the selected urban and rural regions both with (light red and light blue, respectively) and without (dark red and dark blue, respectively) interventions. Results are shown as the mean ± standard deviation over 10 repetitions of each simulation. The incidence rate is calculated separately for each region.}
 \label{fig:ur_int}
\end{figure}

Comparing the effect of interventions on urban and rural regions, we found that interventions have a similar impact on the urban and rural regions, reducing and delaying the peak of infections in both regions (Figure \ref{fig:ur_int}). Further, both with and without interventions, the incidence rate in the urban region peaks earlier than that in the rural region. However, it should be noted that while the incidence rates show similar patterns, the case numbers in the rural region are much lower than those in the urban region, both with and without interventions.

\subsection*{The impact of interventions may be predicted on a national scale}

Finally, to demonstrate how EpiGeoPop and Epiabm can be used together to efficiently construct, run, and visualise epidemic simulations, we constructed a model for the spread of COVID-19 in New Zealand. We compare the effects of different strength interventions to illustrate the range of research that may be conducted with these tools. 

We select New Zealand for this case study due to their rapid implementation of a series of strict intervention strategies which successfully controlled the spread of COVID-19 in the early stages of the pandemic \cite{barrett_islands_nodate}. We sought to model the timing of interventions actually applied from March to May 2020, although do not undertake detailed parameterisation of the model such that the actual dynamics of COVID-19  are captured. However, we do confirm that the reproduction number ($\mathcal{R}$) during the initial simulation period is sensible and within the range of COVID-19 variants (see Supplementary Information Section 1.6 for details of this procedure). 

We utilised the travel functionality implemented in the new version of Epiabm to introduce one new COVID-19 case to New Zealand each day. The parameter values for non-pharmaceutical interventions were based on instructions from the New Zealand government as far as possible. Details of the parameters used are provided in the Supplementary Information (Section 1.4). 

\begin{figure}
\centering
\includegraphics[width=0.5\textwidth]{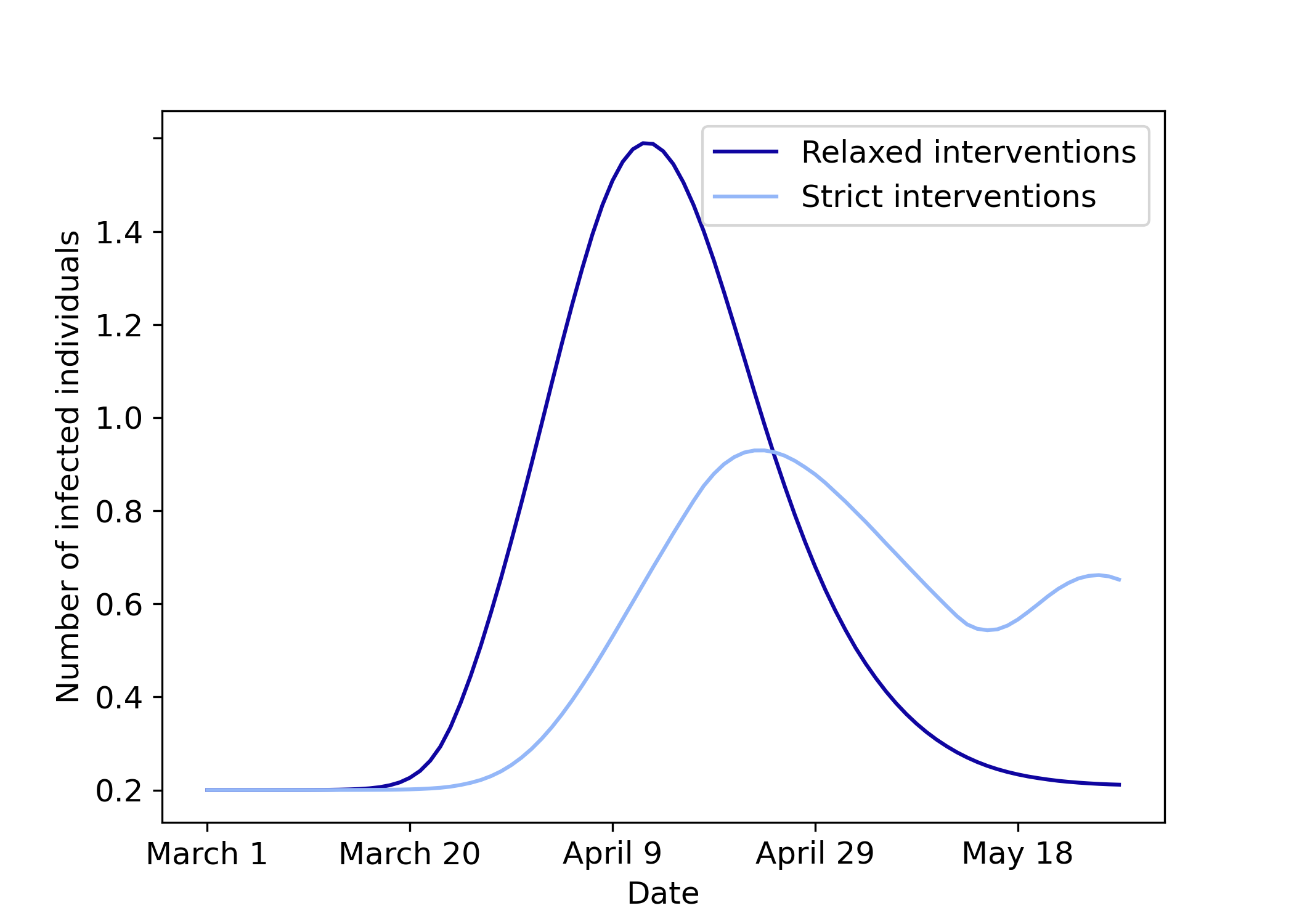}
\caption{Number of infected individuals over time with relaxed and strict interventions. The timings of these intervention followed those conducted by the New Zealand government. The effectiveness of case isolation and household quarantine was impaired in the relaxed intervention.}
 \label{fig:NZ}
\end{figure}

We compared the number of infections under strict and more relaxed interventions, with the effectiveness of case isolation and household quarantine impaired in the relaxed scenario. The number of infections peaked 12 days earlier and at a higher number of cases in the relaxed scenario compared to the stricter setting; under less strict interventions infections peaked at an 91\% higher value than the first peak under the more strict interventions, corresponding to 66.1k additional cases at the peak (Figure \ref{fig:NZ}). The second wave under the strict interventions coincides with the reopening of schools, workplaces, and outdoor spaces. Under the relaxed scenario the number of infections declined more rapidly while under the more strict interventions a second wave of infections was observed. Finally, the relaxed scenario lead to a greater number of deaths over the course of the simulation than under the strict interventions; at day 90, there were 15k more total deaths under related interventions than under strict ones. 

\section*{Discussion}

Given the role mathematical models played in informing government policy in the recent COVID-19 pandemic, accurate simulation of the spread of disease under hypothetical interventions is crucial for pandemic preparedness. Attention has been paid to the reproducibility of the software used to simulate epidemics but ensuring a transparent, reproducible origin of inputs is also crucial. Here, we provide EpiGeoPop, a tool which prepares spatial inputs -- of central importance to ABMS -- in a transparent and reproducible manner.

EpiGeoPop offers a Python workflow for generating and visualizing population input files based on population density data. EpiGeoPop generates population input files and parameters relating to age distribution across the population, facilitating the application of ABMs to real world countries and practical research questions. This interfaces with Epiabm \cite{gallagher_epidemiological_2022}, an alternative implementation of CovidSim \cite{ferguson_report_2020} prioritising good software engineering practices, which we extended to simulate a range of pandemic interventions. 

Combining EpiGeoPop with the updated version of Epiabm, we demonstrated the importance of including spatial detail in epidemiological models, showing how incidence rates may differ according to regional density and the impact of interventions on high density regions compared to lower density ones. These results provide a basis for future work investigating how spatial patterns influence the spread of disease and the response to intervention. A number of recent studies have sought to elucidate the impact of population density on the spread of COVID-19, reaching contradictory conclusions regarding whether population density did (e.g., \cite{pascoal_population_2022, afshordi_diverse_2022, arbel_population_2022}) or did not (e.g., \cite{hamidi_does_2020, khavarian-garmsir_are_2021}) influence the spread of COVID-19. However, since these studies have been retrospective in nature, looking for associations between population density and the number of infections, hospitalisations, and deaths, the results are heavily influenced by the interventions in place in these regions and other demographic and socioeconomic characteristics \cite{udell_clinical_2022}. In order to gain insight into the role of population density to aid preparation for future pandemics, predictive studies modelling the impact of a range of interventions in specific geospatial contexts will be essential. 

As more comprehensive datasets become available, a trend that is already apparent following the COVID-19 pandemic, EpiGeoPop could also be leveraged to model the role of major transport hubs in disease spread as well as the effect of movement of people within and between geographic regions. In this way, EpiGeoPop represents a flexible tool for generating spatially accurate input files for investigating the spread of disease, which can be built upon to be of continued use to the field. By automating population configuration, EpiGeoPop helps streamline the process of obtaining and formatting the large amounts of information needed by ABMs, and we hope that its public availability will promote standardisation of input formats within epidemiological modelling.

\section*{Data Availability}
PyEpiabm is available open source at \url{https://github.com/SABS-R3-Epidemiology/epiabm}. EpiGeoPop is available open source at \url{https://github.com/SABS-R3-Epidemiology/EpiGeoPop}.

EpiGeoPop uses population density data from \url{https://jeodpp.jrc.ec.europa.eu/ftp/jrc-opendata/GHSL/GHS_POP_MT_GLOBE_R2019A/GHS_POP_E2015_GLOBE_R2019A_4326_30ss/V1-0/}, regional boundaries from \url{https://www.naturalearthdata.com/downloads/10m-cultural-vectors/}, and age distribution data from \url{https://population.un.org/wpp/Download/SpecialAggregates/EconomicTrading/}.

Details regarding the Luxembourg simulations can be found at \url{https://github.com/SABS-R3-Epidemiology/epiabm/tree/main/python_examples/luxembourg_example}.
Details regarding the New Zealand simulatios can be found at \url{https://github.com/SABS-R3-Epidemiology/epiabm/tree/main/python_examples/NZ_example}.
Information on the timing of restrictions in New Zealand was found at \url{https://covid19.govt.nz/about-our-covid-19-response/history-of-the-covid-19-alert-system/}

\bibliography{sample}

\section*{Acknowledgements}

The authors would like to acknowledge the use of the University of Oxford Advanced Research Computing (ARC) facility in carrying out this work. \url{http://dx.doi.org/10.5281/zenodo.22558}

\section*{Author contributions statement}

L.H., H.C., I.E., N.S., and J.Z. contributed to the development of the the software, generation of results, and writing of the paper. 
K.G., R.C., I.B., D.G. and B.L. conceptualised the work.
D.G. and B.L. managed project administration. 
K.G., R.C., and I.B. supervised the work.
All authors reviewed the manuscript and approved the final version for submission.

\section*{Additional information}

All authors acknowledge funding from the EPSRC CDT in Sustainable Approaches to Biomedical Science: Responsible and Reproducible Research - SABS:R3 (EP/S024093/1).

%A.L.D. was funded as a Roche Pharmaceutical Research employee.

The authors declare that they have no competing interests.

\newpage

\renewcommand{\thetable}{S\arabic{table}}
\renewcommand{\thefigure}{S\arabic{figure}}
\renewcommand{\theequation}{S\arabic{equation}}
\renewcommand{\thesection}{S\arabic{section}}
\setcounter{figure}{0}  
\setcounter{table}{0}  
\setcounter{equation}{0} 
\setcounter{section}{1}

\section*{Supplementary Information}
\subsection{EpiGeoPop usage}
\label{sec:example}
The code snippet provided below (taken from the GitHub README) outlines how to install and run the EpiGeoPop workflow. The example provided here will generate the relevant output files for Luxembourg.
\\

\begin{verbatim}
# Clone the repository
git clone git@github.com:SABS-R3-Epidemiology/EpiGeoPop.git
cd EpiGeoPop

# Create virtual environment (recommended)
python -m venv venv
source venv/bin/activate

# Install dependencies
pip install -r requirements.txt

# Downlaod the raw data (See data/README.md for more information)
bash prep.sh

# Run the snakemake pipeline
snakemake --cores 1
\end{verbatim}

This example will generate the following output files:
\begin{enumerate}
    \item Population density map,
    \item Input file for the epidemiological simulation,
    \item Json file containing age distributions for the population of interest.
\end{enumerate}

Running the Snakemake pipeline for Luxembourg on an Apple M2 Pro took 5.779s, demonstrating the efficiency of this tool for generating spatially accurate population input files for any country of interest, a task which previously represented a substantial bottleneck in the utility of ABMs.
\\

The \verb|prep.sh| file will download the relevant data in advance of running the Snakemake pipeline. As such, by changing the web address of the data sets, updated population density data can easily be used as long as the format matches that of the data currently used. The \verb|prep.sh| file  provided in EpiGeoPop uses data at the resolution of 30 arc seconds.

\subsection{The effect of interventions}
\label{interventions}

\begin{figure}[ht]
\centering
\includegraphics[width=\textwidth]{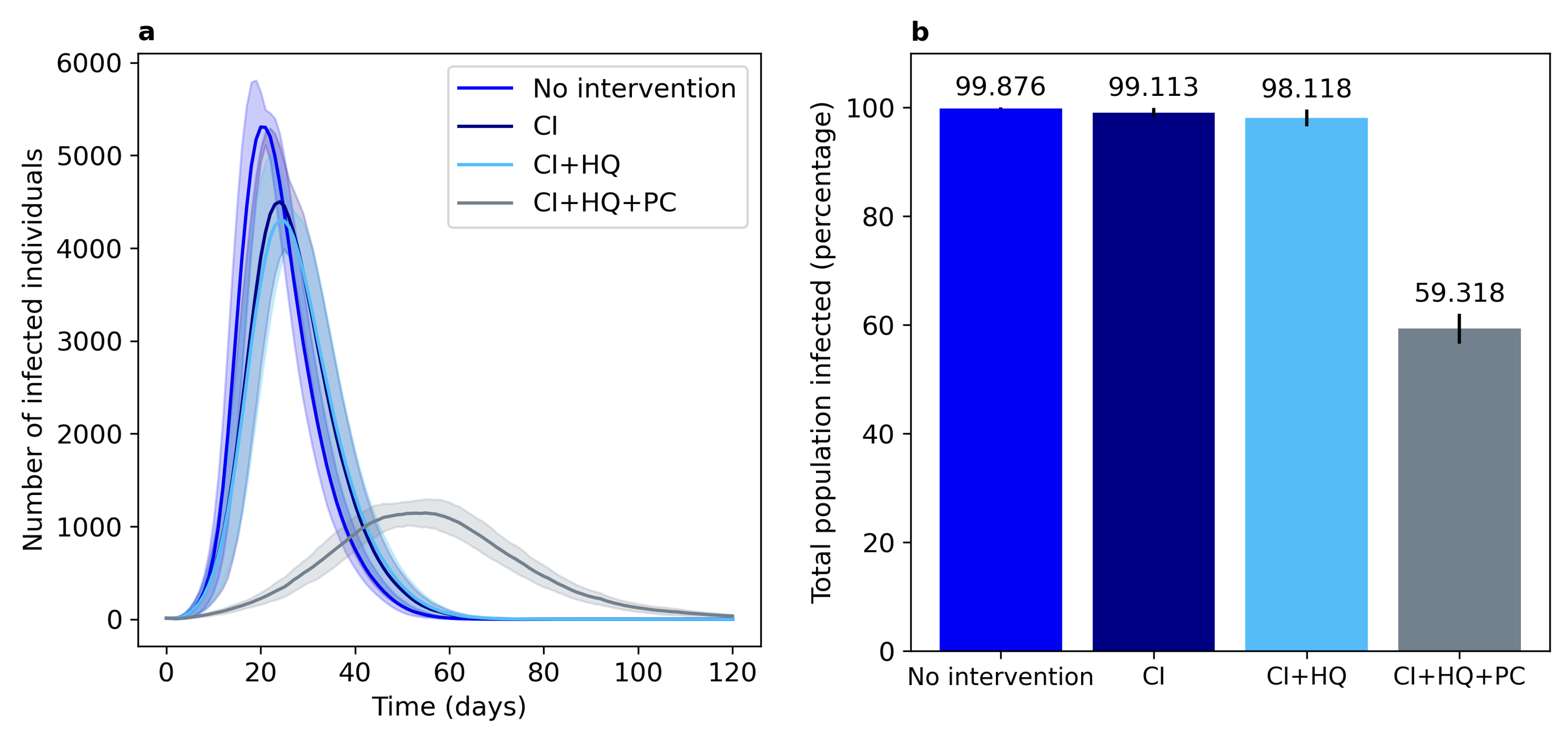}
\caption{Comparison of disease transmission between different number of active interventions. (\textbf{a}) Shows the number of infected individuals over time. (\textbf{b}) Shows the total number of infected individuals over the entire simulation period. Simulations are run on a population of 10,000 individuals spread approximately uniform over a squared 4x4 grid. Without an intervention (dark blue) $97.85 \pm 0.16$ percent of the population will get infected with the highest peak of 5453 infected individuals at day 20. The Case Isolation (CI, blue) intervention showed similar number of infections ($96.69 \pm 0.66$ percent) but the peak of the wave is reduced to 4600 and delayed to day 24. The Household Quarantine (HQ) intervention  as an additional intervention on top of CI (light blue, $95.67 \pm 1.23$ percent) slightly decreased the height of the infectious peak to 4408 at day 24. The total number of infections ($56.94 \pm 3.38$ percent), height of the peak (1089), and the day at which this peak is observed (day 48) are substantial decreased and delayed by introducing the Place Closure intervention as additional intervention (grey).}
\label{fig:toy_interventions}
\end{figure}

As an example, we demonstrate the impact of stacking interventions on a model population, beginning with case isolation and then adding household quarantine and subsequently place closure (see \ref{fig:toy_interventions}). A simulation of 100 days captures the number of infections within the model population of 10,000 individuals spread approximately uniformly over a square 4x4 grid. This is a dense population. The interventions decrease and delay the height of the peak of infections, with Place Closure having the strongest effect. Moreover, Place Closure substantially decreases the total number of infections. The relative impact of each intervention is highly dependent on the context of the population and characteristics of the population in question. As such, the greater impact of place closure compared to other interventions in this example will be influenced by the number of places and households per person in the model population. The parameters used for this simulation can be found at \url{https://github.com/SABS-R3-Epidemiology/epiabm/blob/main/python_examples/model_population_example/IntMult_params.json}.

\subsection{Luxembourg simulation details}
\label{lux}
Luxembourg simulations were run using pyEpiabm version 1.1.0. For the initial conditions used in Figures 3, 4, and 5 we seeded five initial infections in the same cell (number 1664) towards the furthest eastern part of Luxembourg near the border with Germany $(6.400^{\circ} N, 49.708^{\circ} W)$. The urban and rural locations selected, Luxembourg city and Nommern, respectively, are both $\approx 180 km$ from this initial infection site as the crow flies, allowing for fair comparisons to be made. 
Simulations were repeated 10 times and results are presented as mean ± standard deviation. \\

For the simulations with interventions, we applied case isolation and household quarantine for the first 90 days and social distancing between days $49-90$. The parameters governing the simulations can be found at \url{https://github.com/SABS-R3-Epidemiology/epiabm/tree/main/python_examples/luxembourg_example}: for simulations without interventions see luxembourg\_parameters.json and for simulations with interventions see luxembourg\_intervention\_parameters.json.
 
\subsection{New Zealand simulation details}
\label{NZ}
In the New Zealand simulation, the parameter values for non-pharmaceutical interventions were based on instructions from the New Zealand government as far as possible. According to the timeline of Alert Level changes, which documents the dates of key events and the duration of the State of National Emergency (\url{https://covid19.govt.nz/about-our-covid-19-response/history-of-the-covid-19-alert-system/}), we extract the following parameters to use in our simulation of strict intervention. \\

The simulation starts on 1\textsuperscript{st} March and lasts for 90 days. From 14\textsuperscript{th} March, international arrivals are required to quarantine in a hotel for two weeks. Non-traveller individuals must self-isolate for two weeks if they are symptomatic for the duration of the simulation. Members of the household of symptomatic individuals are subject to household quarantine for two weeks. Household quarantine is set to have a 75\% compliance rate while that of case isolation is 100\%.  In addition, all schools, workplaces and outdoor spaces are closed from day 25 to day 74. Social distancing is conducted from day 49 to the end of the simulation. \\

Simulations are run on high-performance computing facilities provided by the University of Oxford. Simulations are run with one compute node and one process per node,  256 GB CPU memory. Due to limited access to compute resources, only one repetition was run to provide an qualitative insight into the impact of interventions. \\

The parameters governing the simulations can be found at \url{https://github.com/SABS-R3-Epidemiology/epiabm/tree/main/python_examples/NZ_example}: for simulations with strict interventions see NewZealand\_parameters.json and for simulations with more relaxed interventions see NewZealand\_parameters\_relaxed.json.

\subsection{The balance between household, place, and spatial infections}

The basic reproduction number ($\mathcal{R}0$) in ABMs, and models with multiple levels of mixing more generally, may be challenging to compute as it emerges from the interaction of multiple different transmission mechanisms. In Epiabm the infectiousness of the virus is governed by three parameters, ``spatial\_transmission", ``household\_transmission", and ``place\_transmission", which act on top of the defined infectiousness profile to control the number of individuals infected at each time step through spatial, household, and place pathways, respectively. Translating these parameters to an effective $\mathcal{R}0$ is non-trivial, and a range of different approaches have been implemented in the literature \cite{ferguson_strategies_2006, hunter_open_data_driven_2018, pellis_reproduction_2012}. For example, CovidSim scales the number of spatial infections by the parameter ``local\_beta", which is calculated by considering the number of spatial infections which are needed on top of place and household infections to obtain the target $\mathcal{R}0$ selected by the user as an input parameter. \\

% An open-data-driven agent-based model to simulate infectious disease outbreaks - this paper defines R0 as contacts*probability of infection*duration of infectiousness. number of contacts is pre-determined from simulation then probability of infection event per contact adjusted as necessary.
% Strategies for mitigating influenza pandemic fits model to data to estimate the R0 (and notes the difficulties with different infection pathways).
% Reproduction numbers for epidemic models with households and other social structures. I. Definition and calculation of R0 - this paper uses branching processes and notes difficulties when have multiple levels of mixing

\begin{figure}[h]
\centering
\includegraphics[scale =0.8]{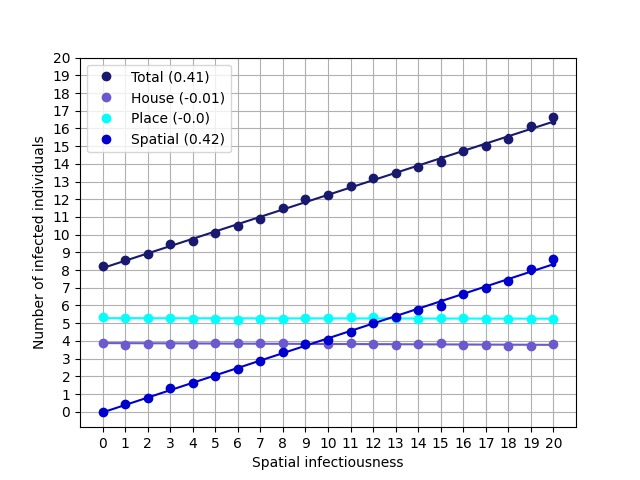}
\caption{Household, place, spatial, and total infections for varying spatial infectiousness parameter values. Simulations were run on a population of 10,000 uniformly distributed over 200 cells for 21 days following a single initial infection. The infections events in their household (purple), the places they visit (light blue), spatially (blue), and the total infections (dark blue) of this single individual over a period of 21 days are shown. Linear curves were fitted and slopes are indicated between brackets in the figure labels. Spatial infections scale linearly with the spatial infectiousness parameter with a slope of 0.42.} 
\label{fig:spatial_inf}
\end{figure}

We studied the influence of the spatial infections parameter on the infection events that take place in households, places, and spatially by infecting one person in an entirely susceptible population. The code used for this investigation can be found at \url{https://github.com/SABS-R3-Epidemiology/epiabm/tree/study-r0/python_examples/r0_testing_simulation}. The number of infections caused by this initial infection over a period of 21 days is studied for different spatial infectiousness parameter values (Figure \ref{fig:spatial_inf}). Secondary infections are not counted. As such, these investigations provided a measure of $\mathcal{R}0$ as traditionally defined. Household and place infections are not affected by the spatial infectiousness parameter, while spatial infections scale linearly with a slope of 0.42. As such, the total number of infections scales in the same way as spatial infections alone, but shifted to a higher value according to the number of household and place infections. \\

% this reference also shows households but its in german...
% chrome-extension://efaidnbmnnnibpcajpcglclefindmkaj/https://www.rki.de/DE/Content/Infekt/EpidBull/Archiv/2020/Ausgaben/38_20.pdf?__blob=publicationFile

The ``spatial\_infectiousness" parameter here, and the ``local\_beta" parameter in CovidSim, only scale spatial infections. However, these spatial infections include all infections between cells, and are thus intended to capture a proportion of infections in places and in visits to other households. The concept of place infections in particular therefore needs to be more cautiously interpreted as only representing infections through places within the person's immediate local area (their cell). Furthermore, Figure \ref{fig:spatial_inf} shows that place infections dominate over the other transmission pathways for spatial infection values below 12. However, a number of previously published analyses based on contact tracing data and genomic surveillance \cite{household_dominant, 10.1093/cid/ciab100} indicate that household infections dominate.\\

Addressing the substantial challenges associated with implementing $\mathcal{R}_{0}$ in agent based models is beyond the scope of this paper. To ensure our simulations fall within the reasonable range for $\mathcal{R}_{0}$, we use an approximate value for ``spatial\_transmission" which results in an initial time-dependent reproduction number ($\mathcal{R}_{t}$) of approximately 3 across all simulations. $\mathcal{R}_{t}$ is the expected number of infections generated by someone infected at time t over the course of their infectious period \cite{creswell_heterogeneity_2022, thompson_key_2020}. We determine this by running inference on the infection curves for each result to calculate a rolling estimate of $\mathcal{R}_
{t}$ \cite{creswell_heterogeneity_2022}. 

\subsection{\texorpdfstring{Inference of $\mathcal{R}_{t}$ using branchpro}{Inference of \emph{R{t}} using branchpro}}
\label{inference}
\begin{figure}[h]
\centering
\includegraphics[width=\textwidth]{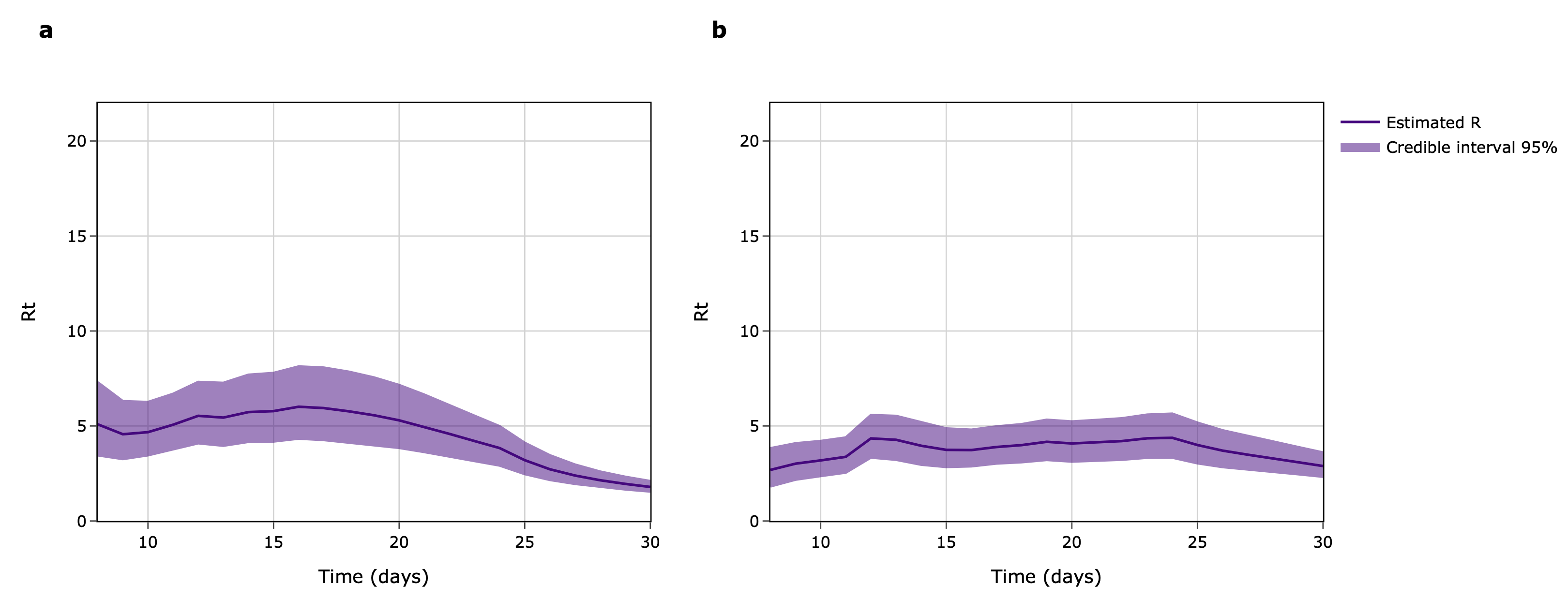}
\caption{Inference of $\mathcal{R}_{t}$ under the strict and more relaxed interventions in New Zealand. \textbf{(a)} $\mathcal{R}_{t}$ profile for the relaxed interventions scenario. \textbf{(b)} $\mathcal{R}_{t}$ profile for the strict interventions scenario. In both plots the solid purple line indicates the mean estimate of $\mathcal{R}_{t}$ and the shaded purple area represents the 95\% central credible interval of the  $\mathcal{R}_{t}$ posterior.}
\label{fig:r0_inf}
\end{figure}

We use branchpro to infer the $\mathcal{R}_{t}$ profile over the first 30 days under both the strict and relaxed interventions in New Zealand (Figure \ref{fig:r0_inf}). We use serial intervals tailored to COVID-19 and the following parameters for the inference: $\tau$, the rolling window over which $\mathcal{R}_{t}$ is averaged, was set at 7 days;  $\epsilon$ was set to 1, meaning the transmission risk of an imported case is deemed equal to local ones; and $\alpha$ and $\beta$, the shape and rate parameters of the gamma prior, were set to 1 and 0.2, respectively. \\

We take the mean inferred $\mathcal{R}_{t}$ over the first 7 days as an estimate of $\mathcal{R}_{0}$ over each simulation. Under the strict interventions we obtain an estimate of $\mathcal{R}_{0}$ of 3.55 ± 1.02, while under the more relaxed interventions we estimate $\mathcal{R}_{0}$ to be 5.16 ± 1.62.

\end{document}